\documentstyle{llncs}

\input epsf

\title{Performance Evaluation for Clustering Algorithms in
       Object-Oriented Database Systems}
\author{J\'{e}r\^{o}me Darmont\\Ammar Attoui\\Michel Gourgand}
\institute{Universit\'{e} Blaise Pascal -- Clermont-Ferrand II,
Laboratoire d'Informatique,
Complexe scientifique des C\'{e}zeaux,
63177 Aubi\`{e}re Cedex,
France}

\begin{document}

\maketitle

\begin{abstract}

It is widely acknowledged that good object clustering is critical to the performance
of object-oriented databases. However, object clustering always involves some kind of
overhead for the system. The aim of this paper is to propose a modelling methodology in order
to evaluate the performances of different clustering policies. This methodology has been used
to compare the performances of three clustering algorithms found in the literature (Cactis, CK
and ORION) that we considered representative of the current research in the field of object
clustering. The actual performance evaluation was performed using simulation. Simulation
experiments we performed showed that the Cactis algorithm is better than the ORION
algorithm and that the CK algorithm totally outperforms both other algorithms in terms of
response time and clustering overhead.


\end{abstract}

\section{Introduction}

        Clustering is a technique that is widely used to improve the performances of
conventional Database Management Systems (DBMSs). Clustering means storing
related objects close together on secondary storage so that when one object is
accessed from disk, all its related objects are also brought into memory. Then access
to these related objects is a main memory access that is much faster than a disk
access.
        With the arrival of Object-Oriented Databases (OODBs), existing clustering
algorithms (mainly in the field of relational databases) had to be adapted to the
additional semantics (such as inheritance, etc.) introduced by the object-oriented data
models. It appeared that a good object clustering was critical to the performance of
OODBs \cite{TSAN91}.
        The aim of this paper is to propose a methodology in order to compare the
performance of the different clustering strategies that can be implemented in
OODBs. Several methods can be used to evaluate the performances of a DBMS.
Benchmarks generally propose a standard database and a series of operations that
run on this database. Thus, performance measurement directly depends on the
reactions of the tested DBMS. Several benchmarks have been specifically designed
for object-oriented databases, like the Synthetic Benchmark \cite{KIM90}, the
HyperModel Benchmark (\cite{ANDE90}, \cite{BERR91}), the OO1 Benchmark \cite{CATT91} or
the CluB-0 Benchmark \cite{BANC92}. However, some OODB designers or clustering
algorithm authors prefer to use simulation (\cite{CHAN89}, \cite{CHEN91}, \cite{HE93}), because
simulation allows to specifically measure performance improvements due to one or
another clustering policy. \cite{TSAN92} proposes a dual performance evaluation
method, performing simulations that use the database introduced by the CluB-0
Benchmark. One last way to determine the advantages of a given clustering method
is mathematical analysis as it is performed by \cite{CHAB93}. This approach is however
limited because the obtained results are qualitative rather than quantitative and sharp
performance criteria cannot be extracted.
        The aim of the approach we suggest to use is to propose a modelling methodology
that allows to compare OODBs' performances, and especially clustering strategies'
performances. Modelling may lead either to simulation or to the application of exact
analytical methods whenever possible. We applied our methodology to three object-oriented
clustering algorithms that are representative of the current research in the
field of OODBs: Cactis \cite{HUDS89}, CK \cite{CHAN90} and ORION (\cite{BANE87}, \cite{KIM90}).
        The main advantage of our approach opposed to the use of benchmarks is that it
allows, by providing a common environment, to specifically compare clustering
algorithms, in a way that is totally independent of any environment associated with
the DBMSs that implement the clustering algorithms we intend to compare. For
instance, physical storage methods and buffering strategies also influence the
DBMS' global performance.
        Furthermore, our approach also allows to a priori study the behaviour of
algorithms (like CK) that are not implemented in any DBMS. Thus we can compare
their performances to those of already implemented algorithms.
        This paper is organised as follows. We start by presenting the modelling
methodology we used. Section~3 is dedicated to our study: we apply our modelling
methodology to obtain a knowledge model and an action model. Then we present in
Section~4 the three studied clustering algorithms. The simulation results are given in
Section~5. We end this paper with a conclusion and a brief discussion about future
research directions.

\section{Modelling Methodology}

        OODBs are complex systems. Modelling their behaviour may as well be a
complex task. This is the reason why we propose an approach dedicated to the study
of such systems. This modelling approach carry out a model according to an iterative
process \cite{GOUR91}. This process is divided into four phases:

\begin{itemize}
\item {\it Phase 1:} analysis and formalizing of data, this system specification
leads to the design of the {\it knowledge model}; it is a crucial step in the modelling process;
\item {\it Phase 2:} translation of the knowledge model into an {\it action model} using
a formalism allowing its exploitation to provide performance criteria;
\item {\it Phase 3:} exploitation of the action model to provide performance criteria;
\item {\it Phase 4:} results interpretation and decisions about actions to
perform on the system.
\end{itemize}

        The analysis approach of a system in order to model it is performed through
several steps:
\begin{itemize}
\item         decomposition of the system to identify the different levels;
\item         decomposition of the system into three subsystems;
\item    {\it logical subsystem} specification;
\item    {\it physical subsystem} specification;
\item    {\it decision subsystem} specification;
\item         specification of the communications between the three subsystems.
\end{itemize}

{\it Note:} The system analysis must be iterative so that the same level of detail is
achieved for all the subsystems.

\section{Study}

        We present in this section the application of the methodology we introduced in
the previous section to the domain of object-oriented databases. Though we focus on
the efficiency of clustering strategies, we do not make any reference in this section to
any precise clustering algorithm.

\subsection{Knowledge Model}

        We need to describe in our model the execution of transactions on an object-oriented
database. We assimilated those transactions to flows running through a
system and thus designed a knowledge model using the SADT actigrams' formalism.
The domain analysis has been described by an entity-relationship (E/R) model.

\subsubsection{Logical Subsystem}

        The logical subsystem specifies what are the flows that run through the system.
In the case of DBMSs, these flows are transactions flows. The transactions are
described on two levels: first, their type and then the steps of their execution. The
HyperModel Benchmark (\cite{ANDE90}, \cite{BERR91}) provides 20 different types of
transactions. From those 20, we have isolated and used 15 types of transactions.

\begin{itemize}
\item          {\it Name Lookup:} Retrieve a randomly selected object.
\item          {\it Range Lookup:} Fetch all the instances of a given class such that a given
attribute value is in a given range.
\item          {\it Group Lookup:} Given a randomly selected starting object, fetch all its
descendant versions (Group lookup along versions), all its component objects (Group
lookup along configuration) or all its equivalent objects (Group lookup along
equivalencies).
\item          {\it Reference Lookup:} It is a "reverse" group lookup. Given a randomly selected
starting object, fetch either all its ancestor versions (Reference lookup along
versions) or its composite object (Reference lookup along configurations).
\item          {\it Sequential scan:} Fetch all the instances of a given class.
\item          {\it Closure Traversal:} Given a randomly selected starting object, follow one of the
three structural relationships (i.e., version, configuration or equivalence) to a certain
predefined depth; fetch the resulting object; the followed relationship can be either
always the same (Closure traversal along versions, configurations or equivalencies)
or randomly selected (Random closure traversal).
\end{itemize}

        The different steps in the execution of the transactions include the following
operations:
\begin{itemize}
\item         select an object to access,
\item         access to the page number of the disk page containing an object,
\item         read or write a page on disk (i.e., perform an I/O),
\item         access to the attributes of an object,
\item         update an attribute value,
\item         place an object in a disk page.
\end{itemize}

\subsubsection{Physical Subsystem}

        The physical resources that make up the physical subsystem are divided into two
categories: {\it active resources} that perform some task and {\it passive resources} that do
not directly participate into any treatment but are used by the active resources to
perform their operations (Table~\ref{res}).

\begin{table}
\begin{center}
\caption{Active and Passive Resources}
\label{res}
\vspace{2pt}
\begin{tabular}{|c|c|c|c|}
\hline
\multicolumn{2}{|c|}{\bf Active resources} &
\multicolumn{2}{|c|}{\bf Passive resources} \\
\hline
AR1 & User                 & \multicolumn{2}{|c|}{\it Physical passive resources} \\
\hline
AR2 & Transactions Manager & PR1 & Main Processor \\
\hline
AR3 & Object Manager       & PR2 & Main Memory \\
\hline
AR4 & Buffering Manager    & PR3 & I/O Processor and Disk(s) \\
\hline
AR5 & I/O Subsystem        & \multicolumn{2}{|c|}{\it Logical passive resources} \\
\hline
AR6 & Clustering Manager   & PR4 & Scheduler \\
\hline
\end{tabular}
\end{center}
\end{table}

\subsubsection{Decision Subsystem}

        The decision subsystem specifies what are the functioning or supervision rules in
the DBMS. Each decision rule listed below as examples (Table~\ref{rules}) is associated to an
SADT activity and is also a method of an object identified in the domain analysis.

\begin{table}
\begin{center}
\caption{Decision Rules List}
\label{rules}
\vspace{2pt}
\begin{tabular}{|c|c|c|}
\hline
{\bf Rule code} & {\bf Rule designation} & {\bf Method of object} \\
\hline
R1 & Generate transaction & Transaction \\
\hline
R2.1 & Extract object & Transaction \\
\hline
R2.2.1 & Access page number & Object \\
\hline
R2.2.2 & Access page & Page \\
\hline
R2.3 & Perform operation & Attribute \\
\hline
R3 & Perform clustering & Database \\
\hline
\end{tabular}
\end{center}
\end{table}

\subsection{Action Model}

        We first translated our knowledge model in a generic action model.
After
being validated, the generic action model has been instanced for each tested
clustering algorithm.

To implement our action model (in this case, a simulation model), we used the QNAP2
(Queuing Network Analysis Package 2$^{\mbox{\tiny{nd}}}$ generation) software, ver\-sion~9.0. We
selected this simulation language for several reasons:
\begin{itemize}
\item         QNAP2 is a validated simulation tool;
\item         QNAP2 allows the use of an object-oriented approach (since version 6.0);
\item         QNAP2's algorithmic language (derived from PASCAL) allows a relatively easy
implementation of such complex algorithms as clustering algorithms.
\end{itemize}

\begin{figure}[htb]
\epsfxsize=10.5cm
\vspace*{-1.0cm}
\centerline{\epsffile{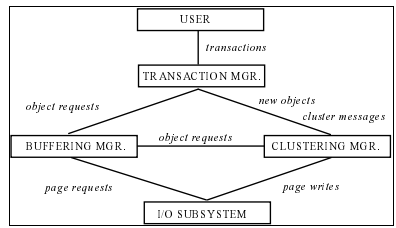}}
\vspace*{-11.0cm}
\caption{QNAP2 Simulation Model Structure}
\label{model}
\end{figure}

Our actual QNAP2 model's main frame is presented in Fig.~\ref{model}.

\section{Studied Clustering Algorithms Presentation}

\subsection{Cactis}

        Cactis \cite{HUDS89} is an object-oriented, multi-user DBMS developed at the University of
Colorado. It is designed to support applications that require rich data modelling
capabilities and the ability to specify functionally-defined data.
        The Cactis clustering algorithm is designed to place objects that are frequently
referenced together into the same block (i.e., page, i.e., I/O unit) on secondary
storage. In order to improve the locality of data references, data is clustered on the
basis of usage patterns. A count of the total number of times each object in the
database is accessed is kept, as well as the number of times each relationship
between objects in the process of attribute evaluation or marking out-of-date is
crossed. Then, the database is periodically reorganised on the basis of this
information. The database is packed into blocks using a greedy algorithm.

{\it Note:} This clustering algorithm is also implemented in the Zeitgeist system \cite{FORD88}.

\subsection{ORION}

        ORION (\cite{BANE87}, \cite{KIM90}) is a series of next-generation database systems that have been prototyped
at MCC (Microelectronics Computer Technology Corp.) as vehicles for research into
the next-generation database architecture and into the integration of programming
languages and databases. ORION has been designed for Artificial Intelligence (AI),
Computer-Aided Design and Manufacturing (CAD/CAM) and Office Information
System (IOS) applications.
        ORION supports only a simple clustering scheme. Instances of the same class are
clustered in the same physical segment (i.e., a number of blocks or pages). Each
class is associated with one single segment.
        Composite objects may also be clustered in multi-classes segments. User
assistance is required to determine which classes should share the segment. The user
can dynamically issue a Cluster message containing a "ListOfClassNames" argument
specifying the classes that are to be placed in the same segment.

\subsection{CK}

        The CK \cite{CHAN90} algorithm (from its authors' names: Chang and Katz) is defined in the
CAD/CAM context. It is not yet implemented in any OODB.
        The CK algorithm is based on a particular inheritance link called instance-to-instance
and inter-objects access frequencies (given by the user at data type creation
time) for each kind of structural relationship (i.e., versions, configurations and
equivalencies). These access frequencies and a computation of the costs of instance-to-instance
inherited attributes give the page where a new object has to be placed. \cite{BULL95}
        The concept of instance-to-instance inheritance is an extension of the classical
inheritance relationship (the IS-A relationship). Instance-to-instance inheritance not
only transfers the existence of attributes from one object to another (like type
inheritance), but moreover the values of these attributes. For example, instance-to-instance
inheritance is important in computer-aided design databases, since a new
version tends to resemble its immediate ancestor. It is useful if a new version can
inherit its attributes values, and more importantly its constraints, from its ancestor.

\section{Simulation Results}

\begin{figure}[htb]
\epsfxsize=10.5cm
\vspace*{-1.0cm}
\centerline{\epsffile{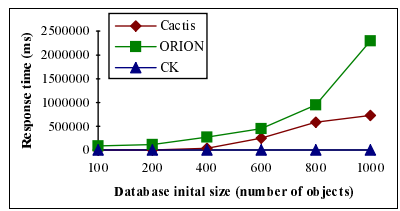}}
\vspace*{-11.5cm}
\caption{Mean Response Time function of Database Initial Size}
\label{resp}
\end{figure}

        Due to a lack of space, we present in this section only a few simulation
results concerning the effects of the database size on the performances.
        Database size directly influence DBMSs' performances, and in particular
clustering algorithms' performances. In this series of simulations, we varied the
database initial size, i.e., the database size before simulation (before new instances
are created).
        Mean response time for each clustering algorithm is given by Fig.~\ref{resp}.
Fig.~\ref{resp}
shows indeed that Cactis is better than ORION (2.5 times better). The CK algorithm
performances are far greater that those of Cactis and ORION (they are
1,100 times better that those of Cactis). This big difference in scale explains
why the results concerning CK do not appear clearly on the graph.

\begin{figure}[htb]
\epsfxsize=10.5cm
\vspace*{-1.0cm}
\centerline{\epsffile{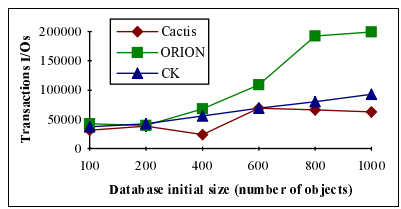}}
\vspace*{-11.5cm}
\caption{Mean Number of Transaction I/O function of Database Initial Size}
\label{strans}
\end{figure}

\begin{figure}[htb]
\epsfxsize=10.5cm
\centerline{\epsffile{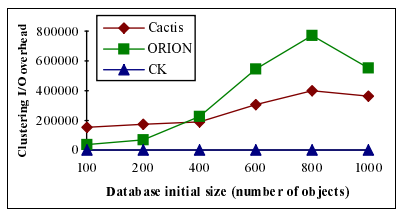}}
\vspace*{-11.5cm}
\caption{Mean Number of Clustering I/O function of Database Initial Size}
\label{clust}
\end{figure}

        These results can be explained by looking at the mean number of I/Os (both
transactions I/Os and clustering I/O overhead) function of the database initial size
(Figs.~\ref{strans} and \ref{clust}). Transactions I/Os giving an idea of how well a clustering
algorithm places the objects, we can deduce from Fig.~\ref{strans} that objects are better
clustered by CK and Cactis than by ORION (2.2 times better for Cactis). Cactis even
appears to be slightly better (1.3 times) than CK.

        The fact that Cactis seems to cluster objects better than CK but shows worse
overall performances can be explained by looking at Fig.~\ref{clust}. It shows
that clustering overhead is 7,000 times greater for Cactis than for CK (clustering
overhead for ORION being 1.4 times greater than for Cactis).

        Such an outstanding performance is due to the true dynamic nature of CK, which
is called only at object creation time and only accesses the object to cluster related
objects, and not to the whole database as Cactis and ORION. Variations in clustering
overhead come from variations in the number of created objects.

\begin{figure}[htb]
\epsfxsize=10.5cm
\vspace*{-1.0cm}
\centerline{\epsffile{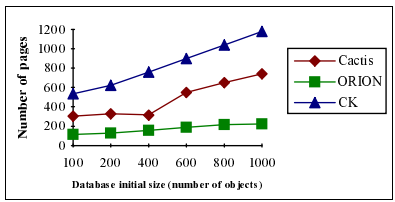}}
\vspace*{-11.5cm}
\caption{Mean Number of Pages function of Database Initial Size}
\label{pages}
\end{figure}

        In terms of disk space, we expected the more sophisticated algorithm to use more
space. Actually, the more a clustering algorithm is complex (i.e., the more it clusters
object according to precise rules), the less a large number of objects are likely to
share the same physical space (either page or segment). The mean number of disk
pages used (Fig.~\ref{pages}), as expected, is higher for the more complex algorithms, i.e.,
CK needs 1.7 times as many pages as Cactis and Cactis needs 2.8 times as many
pages as ORION, for which number of pages increases linearly.

\section{Conclusions}

        It is clear from our simulation experiments that the CK algorithm outperforms
both Cactis and ORION in terms of overall performance. The results we obtained
showed that this is due to both a good clustering capability and to the dynamic
conception of the algorithm that allow an extremely low clustering overhead. Such a
good behaviour is achieved because the CK algorithm is activated only at object
creation time and only accesses the few objects that are related to the newly created
object. Therefore, transactions are never blocked very long during clustering, as they
are when the Cactis or the ORION algorithm is used. (The Cactis and ORION
algorithms have to access all the objects in the database, even several times in the
case of ORION, to reorganise the database; and transactions cannot be run when a
reorganisation occurs.) CK good clustering capability is based on the users' hints that
specify the inter-objects access frequencies for each structural relationship and thus
allows to cluster together objects that are likely to be accessed together.
        Our simulations showed too that Cactis had also a good clustering capability.
This is due to the use of statistics (i.e., objects access frequencies and relationships
use frequencies) that allow to cluster together objects that are actually accessed
together. Though, the Cactis algorithm is still completely outperformed by the CK
algorithm. This is because, when using Cactis, clustering overhead increases very
quickly with the number of objects, thus annihilating any gain achieved from good
clustering capability. However, we have to keep in mind that this algorithm has been
designed to run when the database is idle so that reclustering does not alter the
database performance. Hence, if clustering overhead was not taken into account, the
Cactis algorithm should perform about as well as CK algorithm as long as the
statistics used during the last reorganisation are pertinent.
        In terms of disk space, the ORION algorithm seems the less greedy algorithm.
Then the Cactis algorithm follows, using almost half the number of disk pages
needed by CK to cluster the database. However, when reorganising the database, the
Cactis and ORION algorithms need to build a new set of pages before deleting the
former. Thus they require about twice as much space as our graphs show. Hence,
Cactis and CK are almost equivalent, ORION remaining the less greedy algorithm in
terms of disk space.

        We have presented in this paper a methodology allowing the design of a tool
enabling the a priori study or a posteriori comparison of the performances of
clustering algorithms. This tool may be re-used since it is very easy to instance our
generic action model with other clustering policies than those we chose to study.
This tool may also be modified. It is particularly interesting in future developments
to take into account buffering management strategies because it is mostly the use of
both clustering and buffering techniques rather than clustering techniques alone that
are found in the literature when speaking of performance improvement.
        Our modelling methodology itself may also be re-used to model either another
environment, or to build models designed to test the performances of other
components of an OODB, or even to a priori model the global behaviour of a DBMS
in order to determine some management strategies to use.


\begin{thebibliography}{5}

\bibitem {ANDE90}
T.L. Anderson, A.J. Berre, M. Mallison, H.H. Porter III, B. Scheider:
The HyperModel Benchmark.
International Conference on Extending Database
Technology, Venise, Italie, March 1990

\bibitem {BANC92}
F. Bancilhon, C. Delobel, P. Kanellakis:
Building an Object-Oriented Database System: The Story of O$_{\mbox{\tiny{2}}}$.
Morgan Kaufmann Publishers, 1992

\bibitem {BANE87}
J. Banerjee, H.-T. Chou, J.F. Garza, W. Kim, D. Woelk, N. Ballou, H.-J. Kim:
Data Model Issues for Object-Oriented Applications.
ACM Transactions on Office Information Systems, Vol. 5, No. 1, January 1987

\bibitem {BERR91}
A.J. Berre, T.L. Anderson:
The HyperModel Benchmark for Evaluating Object-Oriented Databases.
In "Object-Oriented Databases with Applications to
CASE, Networks and VLSI CAD", edited by R. Gupta and E. Horowitz, Prentice
Hall Series in Data and Knowledge Base Systems, 1991

\bibitem {BULL95}
F. Bullat:
Regroupement physique d'objets dans les bases de donn\'{e}es.
To appear in ISI, Vol. 3, No. 4, September 1995

\bibitem {CATT91}
R.G.G. Cattell:
An Engineering Database Benchmark.
In "The Benchmark Handbook for Database Transaction Processing Systems", edited by Jim
Gray, Morgan Kaufmann Publishers, 1991

\bibitem {CHAB93}
S. Chabridon, J.-C. Liao, Y. Ma, L. Gruenwald:
Clustering Techniques for Object-Oriented Database Systems.
38$^{\mbox{\tiny{th}}}$ IEEE Computer Society International
Conference, San Francisco, February 1993

\bibitem {CHAN89}
E.E. Chang, R.H. Katz:
Exploiting Inheritance and Structure Semantics for Effective Clustering and Buffering in an Object-Oriented DBMS.
ACM SIGMOD International Conference on Management of Data, Portland, Oregon, June
1989

\bibitem {CHAN90}
E.E. Chang, R.H. Katz:
Inheritance in computer-aided design databases: semantics and implementation issues.
CAD, Vol. 22, No. 8, October 1990

\bibitem {CHEN91}
J.R. Cheng, A.R. Hurson:
Effective clustering of complex objects in object-oriented databases.
ACM SIGMOD International Conference on Management of Data, Denver, Colorado,
May 1991

\bibitem {FORD88}
S. Ford, J. Joseph, D.E. Langworthy, D.F. Lively, G. Pathak, E.R. Perez,
R.W. Peterson, D.M. Sparacin, S.M. Thatte, D.L. Wells, S. Agarwala:
ZEITGEIST: Database Support for Object-Oriented Programming.
2nd International Workshop on Object-Oriented Database Systems,
Bad M\"{u}nster am Stein-Ebernburg, FRG, September 1988

\bibitem {GOUR91}
M. Gourgand, P. Kellert:
Conception d'un Environnement de Mod\'{e}lisation des Syst\`{e}mes de Production.
3$^{\mbox{\tiny{rd}}}$ Industrial Engineering International Congress, Tours, France, March 1991

\bibitem {HE93}
M. He, A.R. Hurson, L.L. Miller, D. Sheth:
An Efficient Storage Protocol for Distributed Object-Oriented Databases.
IEEE Parallel \& Distributed Processing, 1993

\bibitem {HUDS89}
S.E. Hudson, R. King:
Cactis: A Self-Adaptive Concurrent Implementation of an Object-Oriented Database
Management System.
ACM Transactions on Database Systems, Vol. 14, No. 3, September 1989

\bibitem {KIM90}
W. Kim, J.F. Garza, N. Ballou, D. Woelk:
Architecture of the ORION Next-Generation Database System.
IEEE Transactions on Knowledge and Data Engineering, Vol. 2, No. 1, March 1990

\bibitem {TSAN91}
M.M. Tsangaris, J.F. Naughton:
A Stochastic Approach for Clustering in Object Bases.
ACM SIGMOD International Conference on Management of Data,
Denver, Colorado, May 1991

\bibitem {TSAN92}
M.M. Tsangaris, J.F. Naughton:
On the Performance of Object Clustering Techniques.
ACM SIGMOD International Conference on Management
of Data, San Diego, California, June 1992

\end{thebibliography}
\end{document}